\documentclass[prl,twocolumn,aps,showpacs]{revtex4}
\begin{document}
\title{NPPT Bound Entanglement Exists}
\author{Rajiah Simon}
\email{simon@imsc.res.in}
\affiliation{The Institute of Mathematical Sciences, C. I. T. campus, Taramani,
 Chennai - 600113, India}
\date{August 30, 2006}

\begin{abstract} 
Every $d \times d$ bipartite system is
shown to have a large family of undistillable states with
nonpositive partial transpose (NPPT). This family subsumes the
family of conjectured NPPT bound entangled Werner states. In
particular, all one-copy undistillable NPPT Werner states are shown
to be bound entangled.
\end{abstract}
\maketitle

\noindent 
{\bf Introduction:} Maximally entangled bipartite states form an
indispensable resource in several quantum information processing 
situations\,\cite{ekert, cleve, teleport, commit}. In reality, however, readily
available states may be mixed and less than maximally entangled due
to a variety of reasons including influence of the environment. Thus
distillation, the process by which some copies of (nearly) maximally
entangled pure states are extracted from several copies of partially
entangled mixed states using local quantum operations and classical
communication (LOCC), is of singular importance.

Rapid progress has been achieved in this regard, beginning with the
works of Popescu\,\cite{distil1}, Bennett {\it et al.}\,\cite{distil2}, 
Deutsch {\it et al.}\,\cite{distil3}, and Gisin\,\cite{distil4, distil5}. 
An important question that presented itself
at an early stage of this development was this: Can every entangled
state be distilled? The Horodecki family answered this question
affirmatively in a significant particular case: Every inseparable
state of a pair of qubits can, given sufficiently many copies, be
distilled into a singlet\,\cite{allqubits}.

Subsequently they proved a result that applies to arbitrary $m
\times n$ bipartite systems\,\cite{distil-nc}: A state that does not
violate the Peres-Horodecki PPT (positive partial transpose)
criterion\,\cite{peres, peres-horo} can not be distilled. Since
inseparable PPT states can not be distilled, they are said to
possess bound ({\it i.e.}, undistillable) entanglement. The first
examples of such  undistillably entangled PPT states were constructed by
Horodecki\,\cite{range}, and several families of PPT bound-entangled
states have been presented since then. The role PPT
bound-entanglement could play as a quantum resource has also been
studied.

\noindent 
{\bf The Problem of NPPT Bound Entanglement:} Given a
bipartite state $\rho$, let $({\rho}^{T_B})^{\otimes n} =
({\rho}^{\otimes n})^{T_B}$ denote the partial transpose of the
state of $n$ identical copies. Then the necessary and sufficient
condition for distillability of $\rho$ is that the inequality\,\cite{distil-nc}
\begin{equation}
\label{eqn1}
\left\langle{\psi}\right|\left({\rho}^{T_B}\right)^{\otimes
n}\left|{\psi}\right\rangle \ge 0,
\end{equation}
be violated by a Schmidt rank two vector $|\psi\rangle$ in the $n$-copy Hilbert space, 
for some
$n$. Thus in order to be distillable, the state should be NPPT ({\it
i.e.}, the state should have non-positive partial transpose). What
has remained an open problem is this: Are there undistillable NPPT
states? Does undistillability imply PPT\,\cite{open}?

Clearly, violation of the above inequality for a particular $n =
n_0$ implies violation for all $n > n_0$. Given an NPPT state
$\rho$, if it satisfies this inequality for all Schmidt rank two
states, for a particular $n$, we say that the state is $n$-copy
undistillable or, equivalently, that the partial transpose
${\rho}^{T_B}$ is $n$-copy $2$-positive. Thus $\rho$ is
undistillable if ${\rho}^{T_B}$ is $n$-copy $2$-positive for all
$n$.

While Horodecki {\it et al.} pointed out in a subsequent work\,\cite{reduction} that 
NPPT bound entangled states, if they exist in
nature, should be found among the one-copy undistillable NPPT Werner
states\,\cite{wernerstate}, it may be fair to say that this problem 
 attained its present state of fame only when
two leading groups produced independently, and about the same time,  analytical and 
numerical
evidence\,\cite{nppt-divincenzo, nppt-dur} for its existence in the
context of the Werner family of states [More recent numerical attempt
has been undertaken in Ref.\,\cite{nppt-vianna}]. As the analytic
part of the evidence it was shown, for every $n$, that there is a
corresponding range of parameter values over which the NPPT Werner
state remains provably $n$-copy undistillable. But the range itself
becomes rapidly smaller with increasing $n$, and was not proved to
remain nonzero as $n \rightarrow \infty$. It should be added that no
one-copy undistillable NPPT Werner state was shown to be $n$-copy
distillable either. Similar evidence in the multipartite case has
been considered by explicit construction of $n$-copy undistillable
NPPT states, for every $n \ge 1$\,\cite{nppt-som}.

It should be noted for completeness that some systems are known not to support NPPT 
bound entangled states at all. These
include $2 \times n$ dimensional systems\,\cite{allqubits, nppt-dur}
and all bipartite Gaussian states\,\cite{giedke}.

Shor {\it et al.}\,\cite{nppt-shor} have used the conjectured
existence of NPPT bound entanglement to prove nonadditivity of
bipartite distillable entanglement. And Eggeling {\it et al.}\,\cite{nppt-eggeling} 
have related the existence of NPPT bound
entanglement to the connection between the sets of separable
superoperators and PPT-preserving channels. Further, Vollbrecht and
Wolf\,\cite{VW} have shown that an additional resource in the form of
infinitesimal amount of PPT bound-entanglement can render any
one-copy undistillable NPPT state one-copy distillable.

In a more recent paper Watrous\,\cite{nppt-watrous} has constructed,
for every $n \ge 1$, states which are $n$-copy undistillable, and
yet are distillable. This is a surprising and important result, for it
shows that $n$-copy undistillability, even if $n$ is very large,
does not by itself prove  undistillability. The burden this finding places on
numerical evidence for NPPT bound entanglement is evident.

We conclude this brief summary of the present status of the
conjectured existence of NPPT bound entanglement by noting that the
conjecture itself seems to enjoy the confidence of researchers in
quantum information theory, even though the evidence presented so
far has been assessed differently by different 
authors\,\cite{nppt-shor, nppt-eggeling, nppt-watrous}.

In this Letter we prove that any $d \times d$ bipartite system with $d
\ge 3$ has a fairly large family of NPPT states which are
undistillable. As will be seen, this family is much larger than the Werner 
family of conjectured NPPT bound entangled states, but it turns out to be convenient to 
begin our proof with the Werner family.

\noindent 
{\bf Proof of Existence of NPPT Bound Entanglement:} 
 We may define the one-parameter family of  Werner state ${\rho}_{\alpha}$ in $d \times
d$ dimensions through the partial transpose of $\rho_\alpha$:
\begin{equation}
{\rho}^{\rm T_B}_{\alpha} = {\rm Id} - d \alpha P.
\end{equation}
Here $P$ is the projection on
the standard maximally entangled state:
\begin{equation}
P= |\Psi\rangle \langle \Psi |, \,\,\,~~|\Psi\rangle =
\frac{1}{\sqrt{d}} \sum_{k=1}^{d} | k, k \rangle.
\end{equation}
These states as defined are not normalized to unit trace, but this does in 
no way affect our considerations below.

Nonnegativity of ${\rho}_{\alpha}$ forces on $\alpha$ the restriction 
$-1 \leq \alpha \leq 1$. This allowed range for $\alpha$ divides into three
interesting regions\,\cite{nppt-divincenzo, nppt-dur}:
\begin{eqnarray}
-1 \leq \alpha \leq \frac{1}{d}, & \,\,{\rm PPT},\,\,& {\rm separable},
\nonumber \\
\frac{1}{2} < \alpha \leq 1,& \,\,{\rm NPPT},\,\,& {\rm one}{\mathrm -}
{\rm copy\,\,\, distillable}, \nonumber \\
\frac{1}{d} < \alpha \leq \frac{1}{2}, &
\,\,{\rm NPPT},\,\, & {\rm one}{\mathrm -}{\rm copy\,\,\, undistillable}.
\end{eqnarray}
Clearly, it is the last mentioned range which is of interest for the issue on
hand. In this range ${\rho}_{\alpha}$ is NPPT, yet ${{\rho}_{\alpha}}^{T_B}$
is $2$-positive. Since ${\rho}_{\alpha}$ being
NPPT implies that ${\rho}_{\alpha}^{\otimes n}$ is NPPT, 
for all $n$, the issue really is
whether $({{{\rho}_{\alpha}}^{T_B}})^{\otimes n}$ is
$2$-positive as well. We show that it indeed is, for all $n$.

Let us denote by ${{\cal S}^{(1)}}$ the collection of all pure and mixed
states of Schmidt rank (number) $\leq 2$\,\cite{terhal}. This is a
convex set whose extremals are all pure states of Schmidt rank
$1$ or $2$. Let $\theta =
({\theta}_{1},{\theta}_{2},\cdots,{\theta}_{d})$ be a $d$-tuple
of angles, and consider the subgroup of $U(d)$, the
$d$-dimensional unitary group, consisting of diagonal matrices
$U_{\theta}$:
\begin{equation}
{(U_{\theta})}_{kl} =
{\delta}_{kl} e^{i {\theta}_{k}}.
\end{equation}
This is the standard maximal abelian subgroup $U_{A}$ of $U(d)$. Now
average each element $\sigma$ of ${{\cal S}^{(1)}}$ over the local group
$U_{A} \otimes {U}_{A}^{*}$:
\begin{equation}
\sigma \in {\cal S}^{(1)}\,\rightarrow\, {\sigma}^{'} = 
\int d \theta\, U_{\theta} \otimes {U}_{\theta}^{*} 
\,\sigma\, {U}^{\dagger}_{\theta} \otimes
{U}_{\theta}^{T}\,.
\end{equation}
Here $d \theta = d{\theta}_{1}d{\theta}_{2} \cdots d{\theta}_{d}$, with
${\theta}_{k}$ running over the interval $[0,2\pi]$,
independently for every $k$. This group-averaging process
is analogous to the diagonal twirl operation\,\cite{nppt-divincenzo}. The resulting 
images of elements of ${\cal
S}^{(1)}$ constitute a new convex set ${\Omega}^{(1)}$, where the
superscript on ${\cal S}$ and $\Omega$ reflects our intention to
extend these considerations to several copies of the Werner
state.

The reduction in complexity achieved by going
from ${{\cal S}^{(1)}}$ to ${\Omega}^{(1)}$ should be
appreciated. For instance, whereas ${{\cal S}^{(1)}}$ has a
$C{P}^{d-1} \times C{P}^{d-1}$ worth of product states among
its extremals, the product states among the extremals of
${\Omega}^{(1)}$ are precisely $d^2$ in number; these are the
standard computational basis states $|k,l \rangle$. Similarly
the pure states of Schmidt rank = $2$ among the extremals
of ${\Omega}^{(1)}$  necessarily have one of the forms
${\alpha}_1|11\rangle + {\alpha}_2|22\rangle$,
${\beta}_2|22\rangle + {\beta}_3|33\rangle$, and
${\gamma}_1|11\rangle + {\gamma}_3|33\rangle$;
 each of these three sets is a Bloch-sphere in worth. And,
therefore the maximally entangled rank-two states are precisely
three circles worth in number, being of the
form $\frac{1}{\sqrt{2}} (|11\rangle + e^{i {\delta}_{1}}
|22\rangle)$, $\frac{1}{\sqrt{2}} (|22\rangle + e^{i
{\delta}_{2}} |33\rangle)$, or  $\frac{1}{\sqrt{2}} (|33\rangle +
e^{i {\delta}_{3}} |11\rangle)$.

Now $2$-positivity of ${\rho}_{\alpha}^{T_B}$ is equivalent to the demand
that ${\rm Tr}({\rho}^{T_B} \sigma) \geq 0$, for all $\sigma
\in {{\cal S}^{(1)}}$. In view of the $U_{A} \otimes {U}_{A}^{*}$
symmetry of ${\rho}^{T_B}$, this is equivalent to the demand that
${\rm Tr}({\rho}^{T_B} \sigma) \geq 0$, for all $\sigma \in
{\Omega}^{(1)}$ and in view of the convexity of ${\Omega}^{(1)}$,
it is sufficient for the extremals of ${\Omega}^{(1)}$ to meet
this demand. This convexity argment is in essence a recognition of 
the fact that  the real-valued expression ${\rm Tr} (\rho^{T_B}_\alpha sigma)$ is 
linear in $\sigma$.

We thus  find that the minimum of ${\rm Tr}({\rho}^{T_B}
\sigma)$ over ${\Omega}^{(1)}$ equals $(1-2\alpha)$, showing that
${\rho}^{T_B}$ is $2$-positive for all $\alpha \leq \frac{1}{2}$.
The minimum is achieved by $\frac{1}{\sqrt{2}} (|11\rangle +
|22\rangle)$, $\frac{1}{\sqrt{2}} (|22\rangle + |33\rangle)$,
and $\frac{1}{\sqrt{2}} (|33\rangle + |11\rangle)$, and by no
other states.

We may note in passing that the local
symmetry of ${\rho}^{T_B}$ is not $U_{A} \otimes {U}_{A}^{*}$,
but the full group $U(d) \otimes {U(d)}^{*}$. We wish to go
beyond the Werner states later in this Letter, and for this
reason we have based our analysis on the subgroup $U_{A} \otimes
{U}_{A}^{*}$, rather than on the full group.

We now move on to consider ${\rho}_{\alpha} \otimes
{\rho}_{\alpha}$, two copies of the Werner state
${\rho}_{\alpha}$. The set ${{\cal S}}^{(2)}$ is constructed as
the collection of $(d^2 \times d^2)$-dimensional  states of Schmidt number $\leq
2$. From the convex set ${{\cal S}}^{(2)}$ we obtain
${\Omega}^{(2)}$ by averaging each $\sigma \in{{\cal S}}^{(2)}$
over the local group $(U_{A} \otimes {U}_{A}^{*}) \otimes (U_{A}
\otimes {U}_{A}^{*})$. It is to be understood that the first
$(U_{A} \otimes {U}_{A}^{*})$ factor acts on the Hilbert space of
the first copy  and the second on that of the second copy,
independently. 

The extremals of ${\Omega}^{(2)}$ are readily enumerated. The rank-$1$
extremals of ${\Omega}^{(2)}$ are necessarily of the form
$|k,l\rangle_1 \otimes |i,j\rangle_2$. They are $d^4$ in number. The rank
two states are of two types.
\begin{eqnarray}
{\rm Type}\,\,{\rm I}:&&~~ (\alpha |kk\rangle_1 + \beta |ll\rangle_1 )
\otimes |ij\rangle_2, \nonumber \\
         &&~~ |kl\rangle_1 \otimes
(\alpha |ii\rangle_2 + \beta |jj\rangle_2 ); \nonumber \\
{\rm Type}\,\,{\rm II}:&&~~ \alpha |kk\rangle_1 \otimes |ii\rangle_2
+ \beta |ll\rangle_1 \otimes |jj\rangle_2.
\end{eqnarray}
At the risk of sounding repetitive,
we emphasize that this is a complete enumeration of the
extremals of ${\Omega}^{(2)}$.

As with the single copy case $2$-positivity of 
${\rho}_{\alpha} \otimes {\rho}_{\alpha}$ is
equivalent to the demand ${\rm Tr}[({\rho}_{\alpha}^{T_B}
\otimes {\rho}_{\alpha}^{T_B}) \sigma] \geq 0$, for all $\sigma
\in {\Omega}^{(2)}$, which in turn is equivalent to the demand
that this condition be met by all the extremal states of
${\Omega}^{(2)}$. The type-I rank-$2$ states being products
across the two copies, cannot bring out genuinely two-copy
properties if any, and thus we are left with only the type-II
states to examine:
\begin{eqnarray} 
\langle \Psi_{II} |{\rho}_{\alpha}^{T_B} \otimes {\rho}_{\alpha}^{T_B}
| \Psi_{II} \rangle &  \nonumber \\
 = &|\alpha|^2  \langle kk | {\rho}_{\alpha}^{T_B} |kk \rangle
\langle ii | {\rho}_{\alpha}^{T_B} |ii \rangle \nonumber \\
 &+ |\beta|^2 \langle ll | {\rho}_{\alpha}^{T_B} |ll \rangle
\langle jj | {\rho}_{\alpha}^{T_B} |jj \rangle \nonumber \\
 &+ {\alpha}^{*} \beta \langle kk | {\rho}_{\alpha}^{T_B} |ll \rangle
\langle ii | {\rho}_{\alpha}^{T_B} |jj \rangle \nonumber \\
 &+ \alpha {\beta}^{*} \langle ll | {\rho}_{\alpha}^{T_B} |kk \rangle
\langle jj | {\rho}_{\alpha}^{T_B} |ii \rangle.
\end{eqnarray}
Now, $2$-positivity of ${\rho}^{T_B}$ is equivalent to the Schwartz
inequality
\begin{equation}
\langle kk | {\rho}_{\alpha}^{T_B} |ll \rangle \leq
[\langle kk | {\rho}_{\alpha}^{T_B} |kk \rangle \langle ll |
{\rho}_{\alpha}^{T_B} |ll \rangle]^{\frac{1}{2}},
\end{equation}
for rank-1 states. Use of this inequality in Eq.(8) proves
\begin{equation}
\langle \Psi_{II} |{\rho}_{\alpha}^{T_B} \otimes {\rho}_{\alpha}^{T_B}
| \Psi_{II} \rangle \geq 0.
\end{equation}
That is one-copy $2$-positivity of the $(U_{A} \otimes {U}_{A}^{*})$ invariant
operator ${\rho}^{T_B}$ implies its two-copy $2$-positivity. And we have
proved 

\noindent
{\bf Theorem 1}: All one-copy
undistillable Werner states, that is  all ${\rho}_{\alpha}$'s
in the entire range $\frac{1}{d} \le \alpha \leq \frac{1}{2}$, are  
two-copy undistillable.

To move on to the $n$-copy case,
assume that ${\rho}_{\alpha}^{T_B}$ is $(n-1)$-copy $2$-positive.
That is $\langle \psi {({\rho}_{\alpha}^{T_B})}^{\otimes (n-1)}|\psi \rangle 
\geq 0$ for all rank-$2$ states of the $(n-1)$-copy Hilbert space. We wish to
prove that this implies ${({\rho}_{\alpha}^{T_B})}^{\otimes
(n)}$ is $2$-positive.

To this end, form ${\Omega}^{(n)}$ by
averaging each $\sigma \in {\cal S}^{(n)}$ over the local group
${(U_{A} \otimes {U}_{A}^{*})}^{\otimes n}$, with one 
  $(U_{A} \otimes {U}_{A}^{*})$ factor acting on the Hilbert space
of each copy independently. Again, the extremals of
${\Omega}^{(n)}$ consist  of rank-$1$ and rank-$2$ pure states.
The rank-$1$ states are of the form $|i_1,j_1\rangle \otimes
|i_2,j_2\rangle\otimes ....|i_n,j_n\rangle$. That is, these are
tensor products of  computational basis states, one picked from each
copy, and thus are $d^{\,2n}$ in number. 

The rank-$2$ extremal (or pure) states in ${\Omega}^{(n)}$ are of two types,
as in the two-copy case:
\begin{eqnarray} 
{\rm Type}\,\,{\rm I}:&&~~ |\Psi_{I}\rangle  = |\psi \rangle \otimes 
|i,j\rangle,\nonumber\\
{\rm Type}\,\,{\rm II}:&&~~ |\Psi_{II} \rangle  =|{\phi}_{1} \rangle
\otimes |i,i\rangle + |{\phi}_{2} \rangle \otimes |j,j\rangle,
\end{eqnarray}
where $|\psi \rangle$ is a $(n-1)$-copy
rank-$2$ state, and $|{\phi}_{1} \rangle$, $|{\phi}_{2} \rangle$ are
$(n-1)$-copy rank-$1$ states. Since type-I states have a product  
structure across the copies, nonnegative
expectation values of $({{\rho}_{\alpha}^{T_B}})^{\otimes n}$ in respect of 
type-I rank-$2$ states follows directly from the assumed $(n-1)$-copy
$2$-positivity, the additional copy offering nothing new in the type-I case.
 The same $(n-1)$-copy $2$-positivity is equivalent to the
validity of the Schwartz inequality for $(n-1)$-copy rank-$1$ states, and
this in turn implies the nonnegativity of the expectation values
of $({{\rho}_{\alpha}^{T_B}})^{\otimes n}$ for type-II states. We have thus
proved, by induction,

\noindent
{\bf Theorem 2}: The one-copy $2$-positive 
$({{\rho}_{\alpha}^{T_B}})^{\otimes n}$'s in the entire parameter range
$\frac{1}{d} \le \alpha \leq \frac{1}{2}$ are $n$-copy $2$-positive, for all $n$.
That is, these one-copy undistillable NPPT Werner states ${\rho}_{\alpha}$ 
are $n$-copy undistillable,
for all $n$, and hence are bound entangled.

Finally, it should be evident that the only property of ${\rho}_{\alpha}$,
apart from its NPPT and
one-copy $2$-positivity properties, used in our analysis is its 
$(U_{A}\otimes {U}_{A})$ 
invariance or, equivalently, the $(U_{A}\otimes {U}_{A}^{*})$ invariance of its partial 
transpose.  It follows that our conclusions apply
to all states with these properties. That is,

\noindent
{\bf Theorem 3}: Every one-copy undistillable NPPT  state in $d \times d$
dimensions is bound entangled if it possesses
$(U_{A} \otimes {U}_{A}^{})$ symmetry.

{\em This is the main result of this Letter}. It shows that the family of NPPT
bound entangled states in $d \times d$ dimensions, for any $d \geq 3$, 
is much larger than what might have  been  anticipated. This point is worth 
illustrating.

It is easily seen that in $3 \times 3$ dimensions the most general
$(U_{A} \otimes {U}_{A}^{*})$ invariant ${\rho}^{T_B}$ has the form
\begin{eqnarray*}
{\rho}^{T_B}=
\left[
\begin{array}{ccccccccc}
{\rho}_{11} & 0 & 0 & 0 & -z_{12} & 0 &0 & 0 &  -{z}^{*}_{31} \\
0 & {\rho}_{12} & 0 &0 & 0 & 0 & 0 & 0 & 0 \\
0 & 0& {\rho}_{13}& 0 & 0 & 0 & 0 & 0 & 0 \\
0 & 0 & 0 & {\rho}_{21}& 0 & 0 & 0 & 0 & 0 \\
{-z}^{*}_{12}& 0 & 0 & 0 & {\rho}_{22} & 0 & 0 & 0 & -z_{23}\\
0 & 0 & 0 & 0 & 0 & {\rho}_{23} & 0 & 0 & 0 \\
0 & 0 & 0 &0 & 0 & 0 & {\rho}_{31} & 0 & 0 \\
0 & 0 & 0 & 0 & 0 & 0 & 0 & {\rho}_{32} & 0 \\
{-z}_{31} & 0 & 0 & 0 & {-z}^{*}_{23} &
0 & 0& 0& {\rho}_{33} 
\end{array} \right] 
\end{eqnarray*}
Clearly, $\rho$ has to be positive semidefinite
in order to be a valid density matrix.
This demand is equivalent to the conditions: $(i)$
 all the diagonal elements
${\rho}_{ij} \geq 0$, and ${\rho}_{ij} {\rho}_{ji} \geq {|z_{ij}|}^2$
for all $i \neq j$. The NPPT requirement demands that the $3 \times 3$
submatrix
\begin{eqnarray*}
\left[ \begin{array}{ccc}
{\rho}_{11} &  -z_{12} & -{z}^{*}_{31} \\
{-z}^{*}_{12} & {\rho}_{22} &  -z_{23}\\
{-z}_{31} & {-z}^{*}_{23} &{\rho}_{33} 
\end{array} \right]  
\end{eqnarray*}
should be nonpositive, and the $2$-positivity demand is equivalent to the
three inequalities ${\rho}_{11}{\rho}_{22} \geq {|z_{12}|}^2$,
${\rho}_{22}{\rho}_{33}\geq {|z_{23}|}^2$, and 
${\rho}_{33}{\rho}_{11}\geq {|z_{31}|}^2$. 
The NPPT demand and the $2$-positivity demands thus 
 involve only the six parameters ${\rho}_{11}$,
${\rho}_{22}$, ${\rho}_{33}$ and $z_{12}$, $z_{23}$, $z_{31}$.
The phases of the complex z-parameters can be tuned by (local)
change of phases of the basis vectors on the $A$ and $B$ sides
(to be precise it is sufficient to carry out the changes on one side only),
but the argument of $z_{12}$ $z_{23}$ $z_{31}$ is
invariant under such gauge transformations.

Thus our family of NPPT bound entangled states in $3 \times 3$ involves, when 
normalized to unit trace,  $12$
parameters;  eight coming from the diagonals ${\rho}_{ij}$, three coming from the
magnitudes of the $z$-parameters, and one gauge-invariant phase. These are
canonical parameters, and do not take into consideration parameters arising
from local unitary transformations.

\noindent
{\bf Acknowledgement}: I would like to thank Sibasish Ghosh,  Solomon Ivan, 
Guruprasad Kar, and Anirban Roy for many discussions on the problem of NPPT bound 
entanglement.

\end{document}